\documentclass[12pt]{cpl}
\usepackage{supertabular}
\newif\ifpdf
\ifx\pdfoutput\undefined
\pdffalse 
\else
\pdfoutput=1 
\pdftrue
\fi

\ifpdf
\usepackage[pdftex]{graphicx}
\DeclareGraphicsExtensions{.jpg,.pdf,.mps,.png,.tif}
\else
\usepackage{lscape,graphicx}
\fi

\usepackage{amsfonts}

\newcommand{\JCPtoCPL}[4]{{#1} {#2} ({#4}) {#3}.}

\newcommand{\Ref}[4]{\JCPtoCPL{#1}{#2}{#3}{#4}}

\newcommand{\jcp}[3]{\Ref{J. Chem. Phys}{#1}{#2}{#3}}
\newcommand{\jpc}[3]{\Ref{J. Phys. Chem.}{#1}{#2}{#3}}
\newcommand{\cpl}[3]{\Ref{Chem. Phys. Lett.}{#1}{#2}{#3}}

\newcommand{\molphys}[3]{\Ref{Mol. Phys.}{#1}{#2}{#3}}

\newcommand{\ijqc}[3]{\Ref{Int. J. Quantum Chem.}{#1}{#2}{#3}}

\newcommand{\physrev}[3]{\Ref{Phys. Rev.}{#1}{#2}{#3}}

\begin{document}
\begin{frontmatter}

\title{The Electronic Mean-Field Configuration Interaction method: II - Improving Guess Geminals}
\author{Patrick Cassam-Chena\"\i}
\address{ 
CNRS-UNSA, Laboratoire de  Math\'ematique J. A. Dieudonn\'e, Facult\'e des\\ 
Sciences, Parc Valrose,  06108 Nice cedex 2, France. cassam@unice.fr\\
and}

\author{Giovanni Granucci}
\address{ 
Dipartimento di Chimica e Chimica Industriale, via Risorgimento 35,\\
56126 Pisa, Italy. granucci@dcci.unipi.it}

\date{}
\maketitle
\begin{abstract}
The purpose of this letter is to show that a rotation of Hartree-Fock canonical orbitals
which minimizes the lowest eigenvalues of a configuration interaction calculation limited
to the mono-excited configurations from a given orbital allows one to construct a better starting guess for the geminal mean field configuration interaction method.
\end{abstract}
\end{frontmatter}
\newpage
\section{Introduction}

In the electronic mean field configuration interaction (EMFCI) method \cite{Cassam06-jcp}, group functions corresponding to a given group structure are optimised, with or without orthogonality constraint, to represent the lowest states of a given symmetry. The expression of an EMFCI wavefunction is not necessarily unique. For example, in the limit case where the group structure consists in an antisymmetrised product of one-electron functions, i.e. the Hartree-Fock (HF) case, it is well-known that any rotation (respectively unitary transformation) does not change  (respectively, changes only the phase of) the HF wave function. Furthermore, if a restricted HF wave function, is considered as a Grassmann's exterior product of geminals:
\begin{equation}
\Psi_{HF}=g_1\wedge g_2\wedge\cdots\wedge g_n
\label{HF1}
\end{equation} 
with $g_i=\psi_i\wedge\overline{\psi}_i$, the $\psi_i$'s being for example the canonical HF orbitals, then an even larger set of transformations leave the RHF wave function invariant
while preserving a geminal product structure. As an example, it is well known that
this wave function can be written as an ``antisymmetrized geminal product'' (AGP) of extreme
type \cite{Bratos65,Coleman00},
\begin{equation}
\Psi_{HF}=g\wedge g\wedge\cdots\wedge g,
\label{HF2}
\end{equation}
where $g=(n!)^{-\frac{1}{n}}(g_1+ g_2+\cdots + g_n)$.

The non-orthogonal geminal mean field configuration interaction (NO-GMFCI) method presented in Ref.\cite{Cassam06-jcp}, which is the simplest case of EMFCI that preserves spin symmetry
for a system containing an even number of electrons, if pushed to convergence (non-orthogonal geminal self-consistent field (NO-GSCF) method) with no basis set truncation, gives a ground
state independent of the particular representation of the guess wave function. This is not the case for the excited states. This is not the case either, for the ground state, if the basis set of geminals used to represent the geminal mean field Hamiltonian is truncated. Such a truncation can be compulsory to deal with large systems.

Exploratory calculations have shown that the AGP form of the HF wave function, Eq.(\ref{HF2}), is not a good guess for a molecular system, as in general the convergence of a NOG-SCF calculation is slower than with the same HF wave function in canonical form, Eq.(\ref{HF1}). It will not be investigated in the present work. Similarly, a random rotation of the canonical orbitals gives a HF wave function of form, Eq.(\ref{HF1}), whose geminals have not proved useful as a guess for subsequent NO-GSCF calculations, in general. Other attempts to use the eigen geminals of the spin-adapted reduced 2-electron Hamiltonian \cite{Karwowski86} or some extremal electron pairs of Kutzelnigg et al. \cite{Kutzelnigg96,Klopper99} have proved unfruitful so far. Our aim in this letter is to demonstrate that the orbitals obtained by a rotation of the canonical HF orbitals which minimizes the energy of some specific, mono-excited states, can provide a better set of guess geminals than those constructed over canonical orbitals.

The letter is organized as follows: In the next section, we explain how we define ``mono-excitation optimized HF orbitals'' (MOHFOs). Then, in Section 3, we compare MOHFOs based geminals with canonical orbital based geminals on simple examples. Finally, we present the conclusions we have drawn from these examples. 
 
\section{Removing HF orbitals arbitrariness}

An EMFCI calculation step consists essentially in a limited configuration interaction (CI) calculation, in the space spanned by the Grassmann product of a set of ``active'', contracted functions, $\Phi_1^0, \ldots ,\Phi_1^{M}$, with a fixed Grassmann
product of spectator, contracted function, $\Psi_2^0\wedge\cdots\wedge\Psi_r^0$. 
That is to say, the solution of an EMFCI step is obtained by the diagonalisation of a Hamiltonian matrix, $H$, whose elements are defined, in Dirac notation, by, $\forall i,j\in \{0,\cdots ,M\}$,

\begin{equation}
H_{i,j}=\langle\Phi_1^i\wedge\Psi_2^0\wedge\cdots\wedge\Psi_r^0\vert H \vert \Phi_1^j\wedge\Psi_2^0\wedge\cdots\wedge\Psi_r^0\rangle,
\label{hamiltonian_matrix1}
\end{equation} 
where $H$ also denotes the Hamiltonian operator in $n$-electron space. In a general EMFCI calculation, the numbers of electrons, $n_i,\ i\in\{1,\ldots ,r\}$,  of the $r$ contracted functions in each Grassmann product have only to satisfy the constraint that their sum equal
the total number of electrons in the system. 

An unrestricted HF (UHF) calculation for an $n$-electron system can be seen as a particular case of EMFCI where
at each step: $r=n$, $n_i=1,\ \forall i\in\{1,\ldots ,n\}$, the spectator, contracted functions are $n-1$ occupied spin-orbitals, $\Psi_i^0=\psi_i,\ \forall i\in\{2,\ldots ,n\}$, and the basis set of active, contracted functions span the orthogonal complement, $\Phi^0_1=\psi_1, \Phi^1_1=\psi_{n+1}, \ldots ,\Phi^{M}_1=\psi_{m}$. Here, $m$ is the number of spin-orbitals, and $M=m-n$. By iterating such a step with the active spin-orbital corresponding to the approximate ground state becoming spectator and a new, occupied spin-orbital becoming active, one can converge towards the UHF solution.

The stationarity of the ground state energy in this algorithm is clearly equivalent to the Brillouin condition \cite{Brillouin33}. 
This algorithm has the advantage that the approximate excited states obtained by diagonalisation of the Hamiltonian matrix,$\sum\limits_{j=1}^M a^k_j\ \Phi_1^j\wedge\Psi_2^0\wedge\cdots\wedge\Psi_n^0 = ( a^k_1\psi_1+a^k_2\psi_{n+1}+\cdots +a^k_M\psi_m) \wedge\psi_2\wedge\cdots\wedge\psi_n,\ k\in\{1,\ldots ,M\}$, have a physical relevance and define a set of virtual spin-orbitals, $ ( a^k_1\psi_1+a^k_2\psi_{n+1}+\cdots +a^k_M\psi_m), \ k\in\{1,\ldots ,M\}$, such that in addition to Brillouin's theorem, the Hamiltonian matrix elements between the mono-excited configurations from $\psi_1$ towards these virtuals are zero. 

In the case of a converged, RHF, $2n$-electron wave function, a CI calculation in a basis set of the form $\psi_1\wedge\overline{\psi}_1\wedge\cdots\wedge\psi_{n-1}\wedge\overline{\psi}_{n-1}\wedge\left( \frac{\psi_{n}\wedge\overline{\psi}_{n+1}+\psi_{n+1}\wedge\overline{\psi}_{n}}{\sqrt{2}} \right), \ldots ,  \psi_1\wedge\overline{\psi}_1\wedge\cdots\wedge\psi_{n-1}\wedge\overline{\psi}_{n-1}\wedge\left( \frac{\psi_{n}\wedge\overline{\psi}_{m}+\psi_{m}\wedge\overline{\psi}_{n}}{\sqrt{2}} \right)$, (the ground state $\psi_1\wedge\overline{\psi}_1\wedge\cdots\wedge\psi_{n}\wedge\overline{\psi}_{n}$ is not coupled to this set by Brillouin's theorem and there is no need to add it), provides eigenfunctions, 
\begin{eqnarray}
\lefteqn{\sum\limits_{j=1}^{m-n} a^k_j\ \psi_1\wedge\overline{\psi}_1\wedge\cdots\wedge\psi_{n-1}\wedge\overline{\psi}_{n-1}\wedge\left( \frac{\psi_{n}\wedge\overline{\psi}_{n+j}+\psi_{n+j}\wedge\overline{\psi}_{n}}{\sqrt{2}} \right) =}\nonumber\\ \lefteqn{\psi_1\wedge\overline{\psi}_1\wedge\cdots\wedge\psi_{n-1}\wedge\overline{\psi}_{n-1}\wedge\left( \frac{\psi_{n}\wedge\left(\sum\limits_{j=1}^{m-n} a^k_j\ \overline{\psi}_{n+j}\right) +\left(\sum\limits_{j=1}^{m-n} a^k_j\ \psi_{n+j} \right) \wedge\overline{\psi}_{n}}{\sqrt{2}} \right)}, 
\end{eqnarray} 
and defines a set of virtual spin-orbitals, $\sum\limits_{j=1}^{m-n} a^k_j\ \psi_{n+j}$, such that in addition to Brillouin's theorem, the Hamiltonian matrix elements between the mono-excited configurations from $\psi_n$ towards these virtuals are zero.

However, the arbitrariness recalled in introduction remains for the choice of the occupied orbitals of a given symmetry. A natural way to remove this arbitrariness following a variational approach is to minimize the second lowest eigenvalue of the CI calculation in the basis set $R(\psi_1)\wedge R(\overline{\psi}_1)\wedge\cdots\wedge R(\psi_{n-1})\wedge R(\overline{\psi}_{n-1})\wedge\left( \frac{R(\psi_{n})\wedge \overline{\psi}_{n+1})+\psi_{n+1}\wedge R(\overline{\psi}_{n})}{\sqrt{2}} \right), \ldots ,  R(\psi_1)\wedge R(\overline{\psi}_1)\wedge\cdots\wedge R(\psi_{n-1})\wedge R(\overline{\psi}_{n-1})\wedge\left( \frac{R(\psi_{n})\wedge\overline{\psi}_{m}+\psi_{m}\wedge R(\overline{\psi}_{n})}{\sqrt{2}} \right)$ with respect to the rotation of the occupied orbitals $R$. Note that, in fact, this rotation can be restricted to the orbitals carrying a given irreducible representation in case of a non trivial spatial symmetry. Then, one can minimize the next lowest eigenvalue, and so on, until the occupied orbitals rotation freedom is exhausted.

This approach is equivalent to the extended Hartree-Fock (EHF) method of Morokuma \cite{Morokuma72}, except that in our case, each CI calculation gives also a complete set of virtuals in one-to-one correspondance with the  $n$-electron eigenfunctions of the CI matrix. However, one can first use Morokuma's method and then perform a CI with the optimized orbitals to obtain also the same final set of virtuals. This is, in fact, a practical way to proceed. This way, not only one removes the arbitrariness in the HF orbitals and obtains what we call ``mono-excitation optimized HF orbitals'' (MOHFOs), but also one can exploit the one-to-one correspondance between virtual orbitals and eigenvalues of a \textit{bona fide} $n$-electron CI matrix to select a reduced set of the former. In the next section, we compare canonical HF orbitals (CHFOs) with MOHFOs to construct guess geminals for the NO-GMFCI method.

\section{Comparison of canonical versus mono-excitation optimised orbitals}

Following the EMFCI line of thought and trying to improve the description of excited states,
it is natural to use MOHFOs rather than CHFOs. Furthermore, it might be necessary to truncate the HF orbital basis set to deal with large systems and the energies of the mono-excited configurations associated with the virtuals can be used to discard those
that are above a given threshold. However, we must ascertain the superiority of MOHFOs on simple examples.

MOHFOs and CHFOs differs both by their occupied and virtual orbitals. The effect of the optimization of both orbital types can be analysed separately. By using a minimal basis set, the improvement due to virtual orbital optimisation can be quenched, because with such a basis set, there is usually only one virtual orbital with the appropriate symmetry to build the relevant, mono-excited configurations. In contrast, the number of variational degrees of freedom available for the optimisation of the occupied orbitals is not affected by the size of the basis set. It only depends upon the number of occupied orbitals of each symmetry. So, we will first investigate the effect of optimising occupied orbitals on several systems with a STO-3G basis set. This has the additional advantage that full CI (FCI) calculations can be performed for comparison.

The simpler example one can think of is arguably LiH. But in this system, the difference between CHFOs and MOHFOs
is extremely small, because the only variational degree of freedom is the rotation between the core orbital and the valence one, and as one expect these two orbitals do not mix. Many other larger systems exhibit the same feature because of symmetry properties of the occupied orbitals. For example,
the $^0\Sigma^+$ HF ground state energy of  Li$_2$ at the internuclear distance of $2$ bohr is -14.395113 a.u. and the first
$^0\Sigma^+$ mono-excited state energy is -14.094230 a.u. by using the HF canonical orbitals.
By optimizing the rotation of the core orbital and the unique valence orbital allowed to mix by symmetry the energy only decreases down to -14.094435 a.u. . Results for this type of systems are not reported further. 

The simpler system where at least two valence orbitals can possibly mix to produce MOHFOs significantly different from CHFOs is BH. It is iso-electronic to Li$_2$ but has only $C_{\infty v}$ symmetry instead of $D_{\infty h}$.
Results for the first $^1\Pi$ state of this system are reported in Table 1 alongside with the low-lying excited states of water.
The first  $^1\Pi$ state of BH correspond to the excitation of the highest $\sigma$ orbital towards the lowest $\pi$ orbitals. As there is only one set of virtual $\pi$ orbitals in the STO-3G basis set, MOHFO and CHFO sets differ only by their occupied orbitals. After
minimizing the lowest $^1\Pi$ excited state energy, it remains possible to minimize the next excited
state energy with respect to a rotation of two occupied $\sigma$ orbitals without changing the already minimized ground state and first excited state energies. However, this proves to have a negligeable lowering effect on the energy value, because the remaining two $\sigma$ orbitals do not mix, having very different natures, one being a core orbital and the other a valence orbital. 
The optimisation of the occupied orbitals of BH reduces the excited state energy by about $5$ mhartree which is more than an order of magnitude larger than in Li$_2$. This energy lowering is preserved after a NO-GSCF calculation, although the MOHFO and CHFO ground states are no longer identical. A linear dependency threshold of $10^{-1}$ was used in the NO-GSCF, that is to say, the geminal functions whose norm after projection in a Gram-Schmidt orthogonalisation process are less than $10^{-1}$, were eliminated from the geminal basis set at each GMFCI step. Such a low tolerance prevented the geminal excited states from loosing their identity as approximate excited states of the whole system during the GMFCI iterations. The LCI  (CI limited to mono-excitations from a given orbital) vertical excitation energy is worse with MOHFOs than with CHFOs compared with FCI, however, for the more accurate NO-GSCF calculation it is significantly improved with MOHFOs with respect to CHFOs.

In the STO-3G basis set, H$_2$O has only one $a_1$ virtual orbital, so, in the optimization of MOHFOs for the first $^1$A$_1$ excited state, (which corresponds to the excitation of the highest $a_1$ occupied orbital to the $a_1$ virtual), there cannot be any effect from a symmetry preserving rotation of virtual orbitals. Therefore again, the difference between MOHFOs and CHFOs depends only upon the rotation of the occupied orbitals. 
There are three occupied $a_1$ molecular orbitals (including one core orbital) that can be rotated by the EHF process. So the variational freedom is equivalent to that of the BH system already studied. And similarly to BH,  after minimizing the lowest $^1$A$_1$ excited state energy, it remains possible to minimize the next excited
state energy with respect to a rotation of two occupied $a_1$ orbitals without changing the ground state and first excited state energies. However, this proves to have a negligeable lowering effect on the energy value. 

Table 1 shows that the first $^1$A$_1$ excited state is lowered by about 5 millihartrees with MOHFOs.
So, a given variational freedom seems to produce an energy lowering always of the same order of magnitude for the optimised level.
In contrast, the lowest $^1$B$_1$ state energy, which correspond to the excitation of the optimised $a_1$ occupied orbital towards the $b_1$ virtual orbital, increases by about 7.5 millihartrees. Therefore, the energy difference between these two states is significantly improved with MOHFOs when compared to the FCI result. 
The $^1$B$_2$ and $^1$A$_2$ states do not correspond to an excitation from the highest occupied $a_1$ orbital, and are essentially unaffected by the optimisation of the latter.

Now let us turn to the effect of optimising virtual orbitals. This can be analysed on a fixed system, (so that the variational freedom for the optimisation of the occupied is fixed), by varying the basis set. The system must be large enough to justify a truncation of the virtual orbital set before the NO-GSCF calculation, since we want to compare the effect of such a truncation with both orbital guesses. For this study, we have chosen the classical example of ozone. The
low-lying singlet and triplet excited states of ozone have been the subject
of numerous theoretical and experimental
investigations \cite{Messmer76,Steinfeld87,Anderson93,Banichevich93,Braunstein95,Borowski95,Naval99,Palmer02,Zhu04,Deppe04,Kowalski04} to quote a few,
because of the importance of this molecule in the upper athmosphere. 

In this study, we will take as references the two most recent theoretical studies which are the only ones predicting the correct order of the lowest singlet energy levels of the Chappuis band \cite{Palmer02,Zhu04}.
We focus attention on the $^1$A$_1$ ground state and the two lowest singlet excited states of ozone. According to Table 5 of \cite{Palmer02}, the next excited state in increasing order of energy is dominated by a diexcited configuration which corresponds to the excitation of HF orbitals pertaining to two different geminal groups. Therefore, it cannot be described properly within the GMFCI approach and would require a more general EMFCI calculation.

The first singlet excited state carries the B$_1$ representation of the $C_{2v}$ symmetry group and is dominated by the excitation of the highest $a_1$ occupied orbital to the lowest $b_1$ virtual orbital. The symmetry preserving EHF process mimimizes the energy of this state by rotating the 6 occupied $a_1$ orbitals and the virtual $b_1$ orbitals. The HF wave function has only one occupied orbital carrying the $a_2$ and the $b_1$ representations. Therefore no symmetry preserving rotation of these occupied orbitals can affect an excited state energy. In contrast, there are four $b_2$ occupied HF orbitals to consider. The second singlet excited state carries the A$_2$ representation and is dominated by the excitation of the highest $b_2$ occupied orbital to the lowest $b_1$ virtual orbital. So, we use the minimization of its energy with respect to a rotation of the four occupied $b_2$ HF orbitals to remove, partly, the arbitrariness of the latter. The virtuals obtained from the first excited state minimization are not reoptimised. (In fact, we have shown that reoptimising them was not significantly improving  the A$_2$).

The results are displayed in Table 2 for the triple zeta valence basis set plus polarization and Rydberg functions (pVTZ+Ryd) used in \cite{Palmer02}, and for the augmented correlation consistent triple zeta valence basis set plus polarization (aug-cc-pvtz) used in \cite{Zhu04}. 
The effect of the EHF optimization is important as seen from the comparison of the two LCI columns. 
In the ``step0'' columns we have performed GMFCI calculations taking successively as active the electron pairs occupying the highest energy geminals made of, respectively, $a_1$ and $b_2$ orbitals, to obtain the vertical excitation energies from the ground state to, respectively, the lowest $^1$B$_1$ and $^1$A$_2$ states. These calculations are ``step 0'' GMFCI because the spectator geminals are the guess geminals constructed either from MOHFOs or CHFOs. 

To appreciate the advantage of MOHFOs over CHFOs, the GMFCI calculations have been performed after truncating the 81 orbital basis functions of the pVTZ+Ryd basis as well as the 165 orbital basis functions of aug-cc-pvtz to only 18 orbitals.
For MOHFOs, these 18 orbitals correspond to the occupied orbitals and the virtuals whose associated the lowest CI energies.
For CHFOs, they are the 18 orbitals corresponding to the dominant component of the selected MOHFOs. Larger sets of virtuals would lead to more accurate results but would reduce the difference between MOHFOs and CHFOs (in the limiting case of the absence of truncation, virtual orbital optimisation would have no effect on calculations which treat equivalently all the virtuals).

All the GMFCI energy differences increase slightly with respect to the LCI calculations because, for symmetry reasons, in the small truncated orbital basis set, there are many more di-excited A$_1$ geminals interacting with the ground state geminal than B$_1$ or A$_2$ geminals interacting with the B$_1$ or A$_2$ excited states considered.  So, the  energy lowering is more important for the ground state than for the excited states. However, the improvement with MOHFOs of the excitation energies, compared to the MCSCF/SDCI values of \cite{Palmer02} observed at the LCI level, remains essentially preserved in step 0 GMFCI calculations. Our results are somewhat further away from the reference results with the larger basis set. However, similar observations can be done for this basis set, in particular the difference between MOHFOs and CHFOs is of the same order of magnitude despite the more drastic truncation of virtuals. So for virtuals, more variational freedom does not imply necessarily better MOHFOs with respect to CHFOs. Note that the right ordering of the energy levels is obtained in all our calculations.

\section*{Conclusion}

The present work shows that, the freedom one has to rotate CHFOs, already exploited by Morokuma and Iwata in the 1970's, can be taken advantage of to obtain improved guess geminals for the GMFCI method.
The orbitals obtained by minimizing the energy of some specific mono-excited states allow one to truncate the initial orbital basis set, used to build the geminal basis functions, and/or to truncate the geminal basis set itself (by means of the linear dependency cutoff), and then, to perform GMFCI calculations of a better accuracy on the targeted spectral transitions than the same GMFCI with the initial CHFOs. 

Of course, this conclusion on the numerical usefulness of MOHFOs for the GMFCI method should be ascertained by investigating more systems. 
However, from a purely theoretical standpoint, MOHFOs are clearly more appealing than CHFOs, since the arbitrariness of the latter can be removed by a simple application of the variational principle which is  natural within the EMFCI philosophy, where a SCF process is seen as succession of MFCI steps.

\section*{Acknowledgements}
We acknowledge Prof. M.H. Palmer for kindly providing us additional details on his work on ozone, and
the french ANR for fundings (Project AHBE). Computations were performed locally and at the SIGAMM Mesocentre.

\newpage
\begin{center}
\begin{supertabular}{c|c|ccccc|}

&\begin{small}state symmetry\end{small}& \begin{small}CHFO/LCI\end{small} & \begin{small}MOHFO/LCI\end{small} & \begin{small}CHFO/GMFCI\end{small} & \begin{small}MOHFO/GMFCI\end{small} & \begin{small}Full CI\end{small}     \\
\hline                                   
BH&$^1\Pi$ & \textit{0.13813}  & \textit{0.13346}  & \textit{0.16102} & \textit{0.15520} & \textit{0.14647} \\
H$_2$O&$^1$B$_2$ & 0.48441  & 0.48441  & 0.49502 & 0.49489 & 0.45746 \\
H$_2$O&$^1$A$_2$ & 0.55629  & 0.55629  & 0.57422 & 0.57417 & 0.54079 \\
H$_2$O&$^1$A$_1$ & \textit{0.64645}  & \textit{.64153}  & \textit{0.63887} & \textit{0.63367} & \textit{0.59786} \\
H$_2$O&$^1$B$_1$ & 0.72412  & 0.73171  & 0.71976 & 0.72724 & 0.69693\\
\hline
\end{supertabular}
\end{center}

\bigskip

\bigskip

Table 1: Vertical excitation energies from ground state in Hartree, (STO-3G calculations; BH equilibrium bond length of 1.2324 angstr\"om from Ref. \cite{Huber79}; H2O equilibrium bond length of 0.958 angstr\"om, bond angle of 104.4776 degrees from Ref. \cite{Hoy79}.\\
Limited configuration interaction (LCI), refers to a CI limited to the mono-excitations from the highest $a_1$ orbital, therefore, the ground state value is the HF energy for both sets of orbitals. GMFCI refers to a crude NO-GSCF calculation with linear dependency threshold set to $10^{-1}$. The states whose energy is italicized are the ones used to optimise the occupied orbitals. 
For water, the energy gap between the $^1$A$_1$ and $^1$B$_1$ excited states, affected by the orbital optimisation, increases with MOHFOs and compares more favorably with the FCI reference. (See main text for the acronyms.)\\

\newpage
\begin{center}
%
\begin{supertabular}{cc|ccccc|}

\hline
basis & sym.& CHFO/LCI & MOHFO/LCI & CHFO/step 0 & MOHFO/step 0 &  Ref.    \\
\hline                                   
pVTZ+Ryd&$^1$B$_1$ & 2.472  & 2.155  & 2.494 & 2.193 & 2.152\footnotemark[1] \\
&$^1$A$_2$ & 2.9105  & 2.645 & 2.916 & 2.659 & 2.268 \footnotemark[1] \\
aug-cc-pvtz&$^1$B$_1$ & 2.721 & 2.364  & 2.681 & 2.399 & 2.073 \footnotemark[2] \\
&$^1$A$_2$ & 3.179  & 2.878 & 3.106 & 2.889 & 2.170 \footnotemark[2] \\
\hline
\end{supertabular}
\end{center}
\footnotetext[1]{\cite{Palmer02}}

\footnotetext[2]{\cite{Zhu04}}
\bigskip

\bigskip

Table 2: Lowest singlet energy levels (Chappuis band) with respect to the ground state for ozone, in eV. The 
pVTZ+Ryd basis set  and corresponding equilibrium geometry are taken from Ref. \cite{Palmer02}: bond length of 1.2924 angstr\"om, bond angle of  116.53 degrees, 3s2p2d Rydberg functions on the central oxygen nucleus only with exponents 0.021, 0.008, 0.0025, 0.017, 0.09, 0.015, 0.008 au, respectively. The aug-cc-pvtz basis set and corresponding equilibrium geometry are taken from Ref. \cite{Zhu04}: bond length of 1.2822 angstr\"om, bond angle of  116.80 degrees. Step 0 refers to the initialisation step of a GSCF calculation where each geminal CI is performed in the mean field of the guess geminals for the spectator groups. The advantage of the MOHFOs, over the CHFOs is preserved in the step 0 GMFCI calculations. (See main text for the acronyms.)


\newpage

\begin{thebibliography}{99}

\bibitem{Cassam06-jcp}
P. Cassam-Chena\"i,  \jcp{124}{194109-194123}{2006}

\bibitem{Bratos65} 
S. Brato\v{z}, P. Durand, \jcp{43}{2670}{1965}

\bibitem{Coleman00}
A.J. Coleman, V.I. Yukalov, \it Reduced density matrices\rm , (Springer-Verlag, New-York, 2000).

\bibitem{Karwowski86}
J. Karwowski, W. Duch, C. Valdemoro, \physrev{A33}{2254}{1986}

\bibitem{Kutzelnigg96}
W. Kutzelnigg, S. Vogtner, \ijqc{60}{235-248}{1996}

\bibitem{Klopper99}
W. Klopper, W. Kutzelnigg, H. Müller, J. Noga, S. Vogtner, \Ref{Topics in Current Chemistry}{203}{21}{1999}

\bibitem{Brillouin33}
L. Brillouin, \Ref{Actualit\'es Sci. Ind.}{71}{159}{1933} 

\bibitem{Morokuma72}
K. Morokuma and S. Iwata, \cpl{16}{192}{1972}

%
%
%
%
%

\bibitem{Messmer76}
R.P. Messmer, D.R. Salahub, \jcp{65}{779}{1976}

\bibitem{Steinfeld87}
J.I. Steinfeld, S.M. Adler-Golden, J.W. Gallagher, \jpc{16}{911}{1987}

\bibitem{Anderson93}
S.M. Anderson, P. Hupalo, K. Mauersberger, \jcp{99}{737}{1993}

\bibitem{Banichevich93}
A. Banichevich, S.D. Peyerimhoff, F. Grein, \cpl{178}{155}{1993}

\bibitem{Braunstein95}
M. Braunstein, R.L. Martin, P.J. Hay, \jcp{102}{3662}{1995}

\bibitem{Borowski95}
P. Borowski, M. F\"ulscher, P.-A. Malmqvist, B.O. Roos,
\cpl{237}{195}{1995}

\bibitem{Naval99}
N. Naval, S. Pal, \jcp{111}{4051}{1999}

\bibitem{Palmer02}
M.H. Palmer, A.D. Nelson, \molphys{100}{3601-3614}{2002}

\bibitem{Zhu04}
H. Zhu, Z.-W. Qu, M. Tashiro, R. Schinke, \cpl{384}{45}{2004}

\bibitem{Deppe04}
S.F. Deppe, U. Wachsmuth, B. Abel, M. Bitterov\'a, S.Y. Grebenshchikov, 
R. Siebert, R. Schinke, \jcp{121}{5191}{2004}

\bibitem{Kowalski04}
K. Kowalski, P. Piecuch, \jcp{120}{1715}{2004}

\bibitem{Huber79}
K. P. Huber and G. Herzberg, ``Molecular spectra and molecular structure, IV. Constants of diatomic molecules'',
(Van Nostrand Reinhold company, New York, 1979).

\bibitem{Hoy79}
A. R. Hoy, P. R. Bunker, \Ref{J. Mol. Struct.}{74}{1-8}{1979}
\end{thebibliography}
\end{document}